\documentclass{aa}
\usepackage{multirow}
\usepackage{epsfig}
\usepackage{subfigure}
\usepackage{txfonts}
\usepackage{tikz}
\usepackage{xcolor}
\usepackage{wasysym}
\usepackage{amssymb}
\usepackage{multicol}
\usepackage{lscape}
\usepackage{hyperref}
\usepackage{graphicx}

\makeatletter
\let\jnl@style=\rm
\def\ref@jnl#1{{\jnl@style#1}}

\def\aj{\ref@jnl{AJ}}                   
\def\araa{\ref@jnl{ARA\&A}}             
\def\apj{\ref@jnl{ApJ}}                 
\def\apjl{\ref@jnl{ApJ}}                
\def\apjs{\ref@jnl{ApJS}}               
\def\ao{\ref@jnl{Appl.~Opt.}}           
\def\apss{\ref@jnl{Ap\&SS}}             
\def\aap{\ref@jnl{A\&A}}                
\def\aapr{\ref@jnl{A\&A~Rev.}}          
\def\aaps{\ref@jnl{A\&AS}}              
\def\azh{\ref@jnl{AZh}}                 
\def\baas{\ref@jnl{BAAS}}               
\def\jrasc{\ref@jnl{JRASC}}             
\def\memras{\ref@jnl{MmRAS}}            
\def\mnras{\ref@jnl{MNRAS}}             
\def\pra{\ref@jnl{Phys.~Rev.~A}}        
\def\prb{\ref@jnl{Phys.~Rev.~B}}        
\def\prc{\ref@jnl{Phys.~Rev.~C}}        
\def\prd{\ref@jnl{Phys.~Rev.~D}}        
\def\pre{\ref@jnl{Phys.~Rev.~E}}        
\def\prl{\ref@jnl{Phys.~Rev.~Lett.}}    
\def\pasp{\ref@jnl{PASP}}               
\def\pasj{\ref@jnl{PASJ}}               
\def\rmxaa{\ref@jnl{RMXAA}}             
\def\qjras{\ref@jnl{QJRAS}}             
\def\skytel{\ref@jnl{S\&T}}             
\def\solphys{\ref@jnl{Sol.~Phys.}}      
\def\sovast{\ref@jnl{Soviet~Ast.}}      
\def\ssr{\ref@jnl{Space~Sci.~Rev.}}     
\def\zap{\ref@jnl{ZAp}}                 
\def\nat{\ref@jnl{Nature}}              
\def\iaucirc{\ref@jnl{IAU~Circ.}}       
\def\aplett{\ref@jnl{Astrophys.~Lett.}} 
\def\apspr{\ref@jnl{Astrophys.~Space~Phys.~Res.}}
\def\bain{\ref@jnl{Bull.~Astron.~Inst.~Netherlands}}
\def\fcp{\ref@jnl{Fund.~Cosmic~Phys.}}  
\def\gca{\ref@jnl{Geochim.~Cosmochim.~Acta}}   
\def\grl{\ref@jnl{Geophys.~Res.~Lett.}} 
\def\jcp{\ref@jnl{J.~Chem.~Phys.}}      
\def\jgr{\ref@jnl{J.~Geophys.~Res.}}    
\def\jqsrt{\ref@jnl{J.~Quant.~Spec.~Radiat.~Transf.}}
\def\memsai{\ref@jnl{Mem.~Soc.~Astron.~Italiana}}
\def\nphysa{\ref@jnl{Nucl.~Phys.~A}}   
\def\physrep{\ref@jnl{Phys.~Rep.}}   
\def\physscr{\ref@jnl{Phys.~Scr}}   
\def\planss{\ref@jnl{Planet.~Space~Sci.}}   
\def\procspie{\ref@jnl{Proc.~SPIE}}   

\makeatother

\newcommand {\apgt} {\ {\raise-.5ex\hbox{$\buildrel>\over\sim$}}\ }
\newcommand {\aplt} {\ {\raise-.5ex\hbox{$\buildrel<\over\sim$}}\ } 
\begin{document}
\title{A deep X-ray view of the bare AGN Ark 120}
\subtitle{VI. Geometry of the hot corona from spectroscopic and polarization signatures}
\author{A. Marinucci
    \inst{1},        
          D. Porquet\inst{2},
          F. Tamborra\inst{3,4},
          S. Bianchi\inst{1},
          V. Braito\inst{5,6},
          A. Lobban\inst{7},
          F. Marin\inst{8},
          G. Matt\inst{1},
          R. Middei\inst{1},
          E. Nardini\inst{9},
          J. Reeves\inst{5}
          \and
          A. Tortosa\inst{10}
          }
   \institute{Dipartimento di Matematica e Fisica, Universit\`a degli Studi Roma Tre, via della Vasca Navale 84, 00146 Roma, Italy
              \email{marinucci@fis.uniroma3.it}
         \and
             Aix-Marseille Universit\'e, CNRS, CNES, LAM (Laboratoire d’Astrophysique de Marseille) UMR 7326, F-13388, Marseille, France
              \and
                Astronomical Institute of the Czech Academy of Sciences, CZ-14100 Prague, Czech Republic
                \and
                 Nicolaus Copernicus Astronomical Center, Polish Academy of Sciences, Bartycka 18, PL-00-716 Warsaw, Poland
                 \and
                 Center for Space Science and Technology, University of Maryland Baltimore County, 1000 Hilltop Circle, Baltimore, MD 21250, USA
                 \and
                 INAF-Osservatorio Astronomico di Brera, Via Bianchi 46 I-23807 Merate (LC), Italy
                 \and
                 Astrophysics Group, School of Physical and Geographical Sciences, Keele University, Keele, Staffordshire, ST5 5BG, UK
                 \and
                 Universit\'e de Strasbourg, CNRS, Observatoire Astronomique de Strasbourg, UMR 7550, F-67000 Strasbourg, France
                 \and
                 Istituto Nazionale di Astrofisica (INAF) – Osservatorio Astrofisico di Arcetri, Largo Enrico Fermi 5, 50125 Firenze, Italy
                 \and
                 INAF-Istituto di Astrofisica e Planetologie Spaziali, Via Fosso del Cavaliere, 00133 Roma, Italy
             }
             

 
  \abstract
   {The spectral shape of the hard X-ray continuum of Active Galactic Nuclei (AGN) can be ascribed to inverse Compton scattering of optical/UV seed photons from the accretion disc by a hot corona of electrons. This physical process produces a polarization signal which is strongly sensitive to the geometry of the scattering medium (i.e. the hot corona) and of the radiation field.}
   {MoCA (Monte Carlo code for Comptonisation in Astrophysics) is a versatile code which allows for different geometries and configurations to be tested for Compton scattering in compact objects.  A single photon approach is considered as well as polarisation and Klein-Nishina effects. In this work, we selected four different geometries for the scattering electrons cloud above the accretion disc, namely an extended slab, an extended spheroid and two compact spheroids.}
   { We discuss the first application of the MoCA  model to reproduce the hard X-ray primary continuum of the bare Seyfert~1 galaxy Ark 120, using different geometries for the hot corona above the accretion disc. The lack of extra-Galactic absorption along the line of sight makes it an excellent target for studying the accretion disc-corona system. We report on the spectral analysis of the simultaneous 2013 and 2014 XMM-{\it Newton} and {\it NuSTAR} observations of the source.   }
   {A general agreement is found between the best fit values of the hot coronal parameters obtained with MoCA and the ones inferred using other Comptonisation codes from the
literature. The expected polarization signal from the best fits with MoCA is then presented and discussed, in view of the launch in 2021 of the Imaging X-ray Polarimetry Explorer (IXPE). }
   {We find that none of the tested geometries for the hot corona (extended slab and extended/compact spheroids) can be statistically preferred, based on spectroscopy solely. In the future, an IXPE observation less than 1 Ms long will clearly distinguish between an extended slab or a spherical hot corona.}

\titlerunning{The geometry of the hot corona in Ark 120}
\authorrunning{A. Marinucci, et al.}
   \maketitle

\section{Introduction}
The spectral shape of the nuclear continuum of AGN, in X-rays, can be well approximated with a cutoff power law, with a photon index in the range $\Gamma=1.6-2.2$ \citep{bianchi09,sp09}. While this parameter is a function of the optical depth and temperature of the scattering medium, the high energy turnover of the power law mainly depends on the temperature \citep{sle76, rl79,st80, lz87, bel99, petr00, phm01}. The current paradigm invokes the presence of a corona of hot electrons above the accretion disc, which efficiently scatter the ultraviolet radiation emitted by the accretion disc up to X-ray wavelenghts \citep[the so called two-phase model:][]{hm91,hmg94}.\\
\indent
{\it NuSTAR} \citep{nustar} is the first satellite capable of focusing hard X-rays above 10 keV and up to 79 keV. With its broad spectral coverage, it has led to a number of works that have shown that optical depths and electron temperatures typically fall in the ranges $\tau=[0.1-4]$ and kT$_e=[10-500]$ keV, depending on the geometry and on the Comptonization model adopted \citep{flk15, flb17, tbm18}. \\
\indent
A second component has been also invoked to reproduce the soft X-ray excess of AGN \citep[i.e. photons in the 0.5-2 keV band in excess of the extrapolation of the hard power-law component:][]{arn85,sgn85}. These models assume a thermal Comptonisation in an optically thick ($\tau=5-50$) and warm (kT$_e=0.1-1$ keV) scattering plasma \citep{ mbz98, pro04, ddj12, jwd12, ppm13, rmb15, pud18}. In this work, however, we focus on Comptonisation from the hot corona only.\\
\indent
So far, X-ray spectroscopy has not been able to determine the geometry in any of the sources observed with {\it NuSTAR}  and included in the catalogs mentioned above (simultaneously with XMM, {\it Chandra}, {\it Suzaku}, {\it Swift}). In fact, while the bet fit parameters may depend on the adopted geometry \citep{tbm18}, the statistical quality of the fit is invariably the same, even for the best quality {\it NuSTAR} AGN spectra. On the other hand, Compton scattering will produce a polarization signal which is strongly dependent on the geometry of the scattering medium. The IXPE \citep[Imaging X-ray Polarimetry Explorer:][]{wro16, wro16b} and eXTP (enhanced X-ray Timing and Polarimetry mission: Zhang et al, 2016) satellites will investigate the polarization properties of the hot coronae in the brightest, unobscured AGN. In particular, {\it IXPE} has been recently selected by NASA as a SMall EXplorer Mission (SMEX) for a launch in 2021 and it will be the first X-ray imaging polarimeter on orbit, operating in the 2-8 keV band.\\
\indent
In this context, a Comptonisation code which includes both special relativity \citep[MoCA: a Monte Carlo code for Comptonisation in Astrophysics;][Middei et al., in prep.]{tmb18} and general relativity effects (Tamborra et al. 2018b, in prep.) as well as polarization has been recently released. Compared to different codes in the literature, such as \textsc{compTT, compPS, nthcomp} \citep{tit94,ps96,zjm96,zds99}, the energy-dependent Klein-Nishina cross section is taken into account \citep[differently from][]{sk10} and  multiple geometries can be tested and investigated \citep[similarly to][]{bkm17}. Moreover, the Monte Carlo approach implies that no a priori limitations in the parameter space is present.\\
\indent
In this work, we constructed spectral models based on {\sc MoCA}, and used it for fitting the 2013 and 2014 XMM and {\it NuSTAR} observations of the broad-line Seyfert 1 galaxy Ark 120, exploring different geometries of the hot corona. We will also use {\sc MoCA} to predict the polarization signal expected in such geometries. The paper is organized as follows: in Sect. 2 we discuss the data reduction procedure and the implementation of the model, while in Sect. 3  we present the spectral analysis. The polarization signals are reported and discussed in Sect.4. We summarize our results in Sect. 5. \\
\indent
Throughout the paper, we adopt the cosmological parameters $H_0=70$ km s$^{-1}$ Mpc$^{-1}$, $\Omega_\Lambda=0.73$ and $\Omega_m=0.27$, i.e. the default ones in \textsc{xspec 12.10.0} \citep{xspec}. Errors correspond to the 90\% confidence level for one interesting parameter ($\Delta\chi^2=2.7$), if not stated otherwise.

\section{Observations and data reduction}
\subsection{XMM-Newton}
Ark 120 has been the target of a deep XMM-{\it Newton} \citep{xmm} observation in 2014,  starting on 2014 March 18 for a total elapsed time of $\sim 650$ ks with the EPIC CCD cameras, the Pn \citep{struder01} and the two MOS \citep{turner01}, operated in small window and thin filter mode. In this work, we only focus on the third XMM orbit \citep[observation 2014c in][]{preem18} since it is the only one simultaneous with {\it NuSTAR}. We also analyse data from the 2013 February 18 pointing (total elapsed time of 130 ks), also simultaneous with {\it NuSTAR}, applying the latest calibration files available on 2018 June. In 2014, the source was observed in a flux state which was a factor of $\sim2$ higher than in 2013. Data from the MOS detectors are not included in our analysis due to the high pile-up and lower statistics of the spectra. The extraction radii and the optimal time cuts for flaring particle background were computed with SAS 16 \citep{gabr04} via an iterative process which leads to a maximization of the Signal-to-Noise Ratio (SNR), similar to the approach described in \citet{pico04}. The resulting optimal extraction radii are 40 arcsec, net exposure times are 89 ks and 92 ks for the 2013 and 2014 observation, respectively. Background spectra were extracted from source-free circular regions with a radius of 50 arcsec. Spectra were then binned in order not to over-sample the instrumental resolution more than a factor of three and to have no less than 30 counts in each background-subtracted spectral channel. Since no significant spectral variability is observed within each observation, we used time averaged spectra \citep[we refer the reader to][for further details]{mmg14, npr16, lpr18,preem18}.\\

\subsection{NuSTAR}
\begin{figure*}
 \epsfig{file=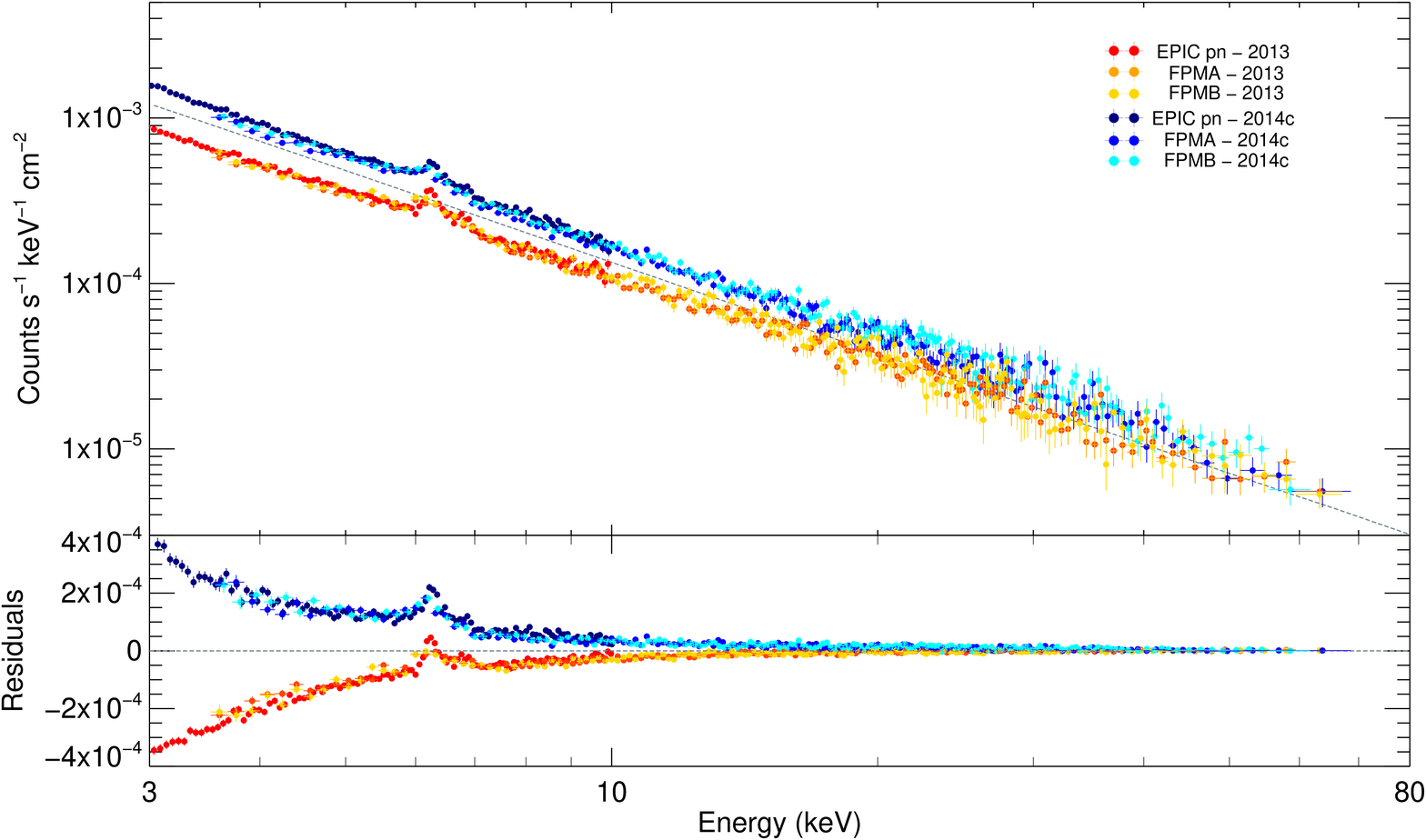, clip=, width=2.0\columnwidth}\\
  \caption{The simultaneous XMM-{\it Newton} and {\it NuSTAR} spectra and residuals to a power law model (in counts s$^{-1}$ keV$^{-1}$ cm$^{-2}$ units) are shown, obtained on February 2013 and March 2014 for Ark 120. Dashed grey line indicate a $\Gamma=1.85$ power law model, with a 2-10 keV flux of $3\times10^{-11}$ erg cm$^{-2}$ s$^{-1}$.}
 \label{plot_spectra}
\end{figure*}

{\it NuSTAR} (Harrison et al. 2013) observed Ark 120 with its two coaligned X-ray telescopes with corresponding Focal Plane Module A (FPMA) and B (FPMB) simultaneously to XMM-{\it Newton} on 2013 February 18 and on 2014 March 22 for a total of $166$ ks and 131 ks of elapsed time, respectively.  The Level 1 data products were processed with the {\it NuSTAR} Data Analysis Software (NuSTARDAS) package (v. 1.8.0). Cleaned event files (level 2 data products) were produced and calibrated using standard filtering criteria with the \textsc{nupipeline} task and the latest calibration files available in the {\it NuSTAR} calibration database (CALDB 20180419). Extraction radii for the source and background spectra were $30$ arcsec and 50 arcsec and the net exposure times for the two observations were 80 ks and 65 ks, respectively.  The two pairs of {\it NuSTAR} spectra were binned in order not to over-sample the instrumental resolution more than a factor of 2.5 and to have a SNR  greater than 5$\sigma$ in each spectral channel. A cross-calibration factor within 3 per cent between the two detectors is found. Broad band (between 3 and 79 keV) data are shown in Fig. \ref{plot_spectra} (top panel) and their residuals to a $\Gamma=1.85$ power law model, with a 2-10 keV flux of $3\times10^{-11}$ erg cm$^{-2}$ s$^{-1}$, indicate a clear spectral variation between the two pointings (bottom panel).
\section{Data analysis}
\subsection{Building the model}
The bare Seyfert galaxy Ark 120 \citep[z=0.0327][]{op77} is an excellent source for studying the accretion disc-corona environment, since no significant extra-Galactic absorption along the line of sight is present, both in the UV and in X-rays \citep{ckb99,vfb04,rpb16}. \\
The source, throughout the years, has shown a variable emission complex in the 6-7 keV band with a broad relativistic Iron K$\alpha$ component in the 2007 {\it Suzaku} spectrum \citep{nfr11} which was not detected in the 2013 XMM+{\it NuSTAR} observation \citep{mmg14}. The neutral core of the 6.4 keV Iron K$\alpha$ fluorescence line was resolved in the 2014 {\it Chandra} High Energy Transmission Gratings spectrum and its Full Width at Half Maximum (FWHM) is consistent with being emitted in the optical Broad Line Region \citep[FWHM=$4700^{+2700}_{-1500}$ km/s:][]{npr16}. Furthermore, the 2014 monitoring performed with {\it Chandra}, XMM and {\it NuSTAR} also led to the discovery of transient Fe K emission from the innermost regions of the accretion disc, both on the red and blue sides of the neutral 6.4 keV line. The full data set is analyzed and discussed in \citet{preem18} and clear variations in the shape of the nuclear continuum were found with respect to the previous 2013 pointing. A steeper continuum was observed in the brighter state, accordingly to the well-known softer when brighter behavior \citep[][]{shem06, rye09,upm16}. The change in the slope of the primary emission and the detection of a high energy cutoff in the 2014 spectra are indicative of a variation of the scattering material configuration. We hereby investigate the geometry and the physical parameters of the hot corona, by using the Monte Carlo Comptonisation model MoCA. { Even though the 2014 monitoring presented flux and spectral changes throughout the four XMM orbits \citep{lpr18, preem18}, we only considered XMM data simultaneous with {\it NuSTAR}, to better characterize the shape of the primary continuum above 10 keV.}\\
\indent
MoCA is a versatile code which allows for different geometries and configurations \citep[we refer the reader to][for a more detailed description of the code]{tmb18}. We selected two different geometries for the scattering electrons cloud above the accretion disc: an extended slab and a sphere. The accretion disc emits a multi-temperature black body from 6 up to 500 gravitational radii (r$_g=GM/c^2$) and both the slab and the sphere cover the whole disc. In all our simulations we considered a black hole mass M$_{\rm bh}=1.5\times 10^8$ M$_{\odot}$ \citep[estimated via reverberation mapping:][]{pet04}, and an accretion rate L$_{\rm Bol}$/L$_{\rm Edd}=10\%$ (which is representative to the values found from the 2013 and 2014 optical to hard X-ray spectral fitting, i.e. 3--7\% (Porquet et al. 2018b). For simplicity, we used spectra integrated over the inclination angle and only Special Relativity effects are considered. We refer the reader to Tamborra et al. (in prep.) for further details on the inclusion of General Relativity effects (which are particularly relevant at radii r$<$6r$_g$) and on the (small) dependence of the emitted spectra on the inclination angle. For the extended slab we assumed a constant height $H=10$ r$_g$ and we simulated spectra in the ranges kT$_e$=20-200 keV and $\tau$=0.1-2.5 adopting 5 keV and 0.1 steps for the temperature and optical depth, respectively. The angle-averaged spectrum is then recorded for each kT$_e$-$\tau$ pair.  
The same procedure is adopted for the spherical geometry, in which we considered a hemisphere covering the disc with temperatures and optical depth in the ranges kT$_e$=20-200 keV and $\tau$=0.1-4.5, due to the different definition of the optical depth in the two geometries (in the sphere it is the radial one, thence equal to the effective one, while in the slab it is the vertical one, therefore smaller than the effective one). As a final step, we considered two scenarios in which the spherical hot corona is more compact, extending from 6 to 100 r$_g$ and from 6 to 20 r$_g$ (see Sect. 3.2.2). The latter is of the order of the coronal size envisaged via microlensing experiments \citep{rm13}, coronal eclipses by clouds \citep{rne11, sanmi13} or reverberation analyses \citep{demarco13, kaf16}. {\sc xspec} readable tables were generated between 0.1 and 700 keV for each configuration.

\subsection{Spectral analysis}
The baseline model adopted for fitting the simultaneous XMM/{\it NuSTAR} observations is the one discussed in \citet[][]{preem18}. It is composed of a primary cutoff power law, relativistic reflection from the accretion disc \citep[modeled with {\sc relxill} in {\sc xspec}, ][and references therein]{gdl14,dgp14}, three Gaussians fixed at 6.4 keV, 6.97 keV and 7.05 keV \citep[see][for a more complete discussion on these components]{npr16} and a warm comptonisation component to reproduce the soft excess \citep[modeled with {\sc compTT}:][assuming a slab geometry]{tit94}. We used a fixed emissivity $\epsilon(r)\propto r^{-3}$ and an inclination angle $i$=30 degrees for the {\sc relxill} component. The whole model is multiplied by a constant, to account for the cross-calibration between the three detectors, while Galactic absorption is modeled with {\sc TBabs}, using a N$_{\rm H}=1.0\times10^{21}$ cm$^{-2}$ \citep{kalberla05}. In {\sc xspec}, the model reads as follows:\\\\
\noindent
{\sc  const$\times$TBabs$\times$(cutoffpl + compTT + relxill + 3$\times$zgauss)}.\\
\noindent
\begin{table}
\begin{center}
\begin{tabular}{cccc}
{\bf Obs. Date} & \multicolumn{2}{c}{\bf Best fit parameter} &{\bf $\chi^2$/d.o.f.} \\\\
\hline
 &\multicolumn{2}{c}{\sc Cutoff power law}& \\
  & $\Gamma$ & E$_{\rm c}$ (keV)&   \\  
 February 2013&   $1.79\pm0.02$  & $>200$ &   569/516\\
 March 2014 & $1.92\pm0.02$  &   $300^{+180}_{-100}$  & 555/499 \\
 &   &    &  \\
   &\multicolumn{2}{c}{\sc NTHcomp}& \\
   &$\Gamma_{\rm Nth}$ & kT$_e$ (keV)&    \\
  February 2013&  $1.80\pm0.02$   &$>40$   & 574/516\\
 March 2014 &  $1.94\pm0.02$    & $155^{+350}_{-55}$ & 560/499 \\
 \hline
      &\multicolumn{2}{c}{\sc CompTT$_{-}$Slab}& \\
   &$\tau$ & kT$_e$ (keV)&    \\
  February 2013&   $0.13^{+0.42}_{-0.10}$  &$>50$  &  568/516\\
 March 2014 &  $0.15^{+0.20}_{-0.05}$ & $150^{+160}_{-75}$    &   555/499\\
      &   &   &  \\
            &\multicolumn{2}{c}{\sc CompTT$_{-}$Sphere}& \\
   &$\tau$ & kT$_e$ (keV)&   \\
  February 2013&   $0.48^{+2.75}_{-0.22}$  &$>40$  &  569/516\\
 March 2014 &  $0.75\pm0.55$ & $125^{+190}_{-65}$    &  555/499 \\
 \hline
   &\multicolumn{2}{c}{\sc MoCA$_{-}$Slab}& \\
   &$\tau$ & kT$_e$ (keV)&  \\
  February 2013&   $0.58^{+0.26}_{-0.08}$  &$110^{+8}_{-17}$  &   566/516\\
 March 2014 &  $0.51^{+0.25}_{-0.20}$ & $103^{+25}_{-20}$    &  552/499 \\
      &   &    &  \\
   &\multicolumn{2}{c}{\sc MoCA$_{-}$Sphere}& \\
   &$\tau$ & kT$_e$ (keV)&    \\
  February 2013&  $3.00^{+0.05}_{-0.55}$  & $57^{+18}_{-2}$ &   561/516\\
 March 2014 & $0.80^{+1.75}_{-0.15}$ &  $120^{+10}_{-40}$   &  551/499 \\
       &   &     &    \\
   &\multicolumn{2}{c}{\sc MoCA$_{-}$Sphere$_{-}$Compact}& \\
   &$\tau$ & kT$_e$ (keV)&    \\
  February 2013&  $1.50^{+0.08}_{-0.30}$  & $95\pm15$&   561/516\\
 March 2014 & $0.35\pm0.05$ &  $>190$   &  552/499 \\
\hline
\end{tabular}
\end{center}
\caption{\label{best_par} Hot coronal best fit parameters are reported, for the five models applied to the data. For the {\sc MoCA$_{-}$Sphere$_{-}$Compact} models, we adopted a spherical hot corona configuration, extended from 6 to 100 r$_g$ and from 6 up to 20 r$_g$ for the 2013 and 2014 observations, respectively.}
\end{table}

\begin{figure*}
 \epsfig{file=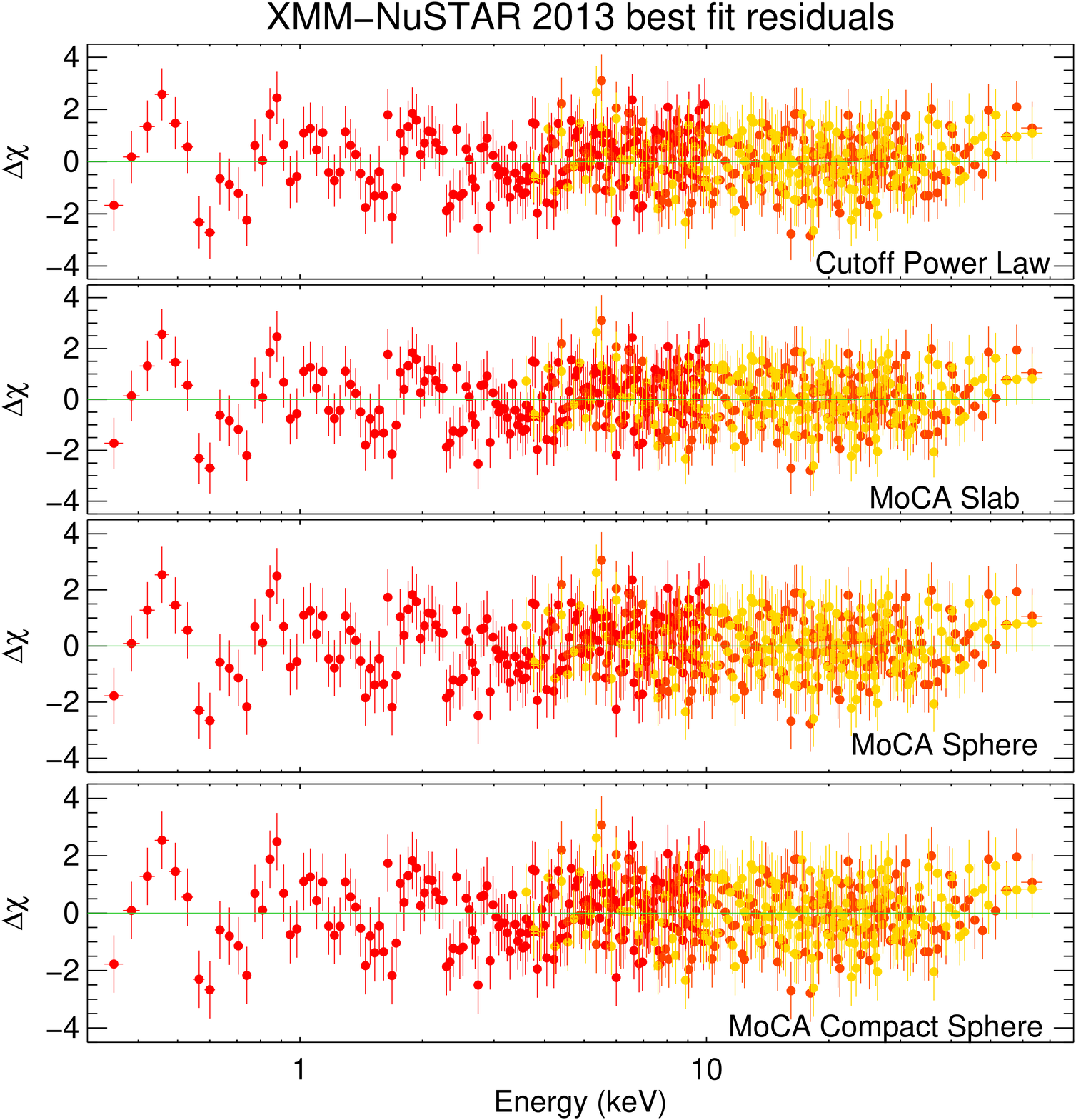, width=0.99\columnwidth}
  \epsfig{file=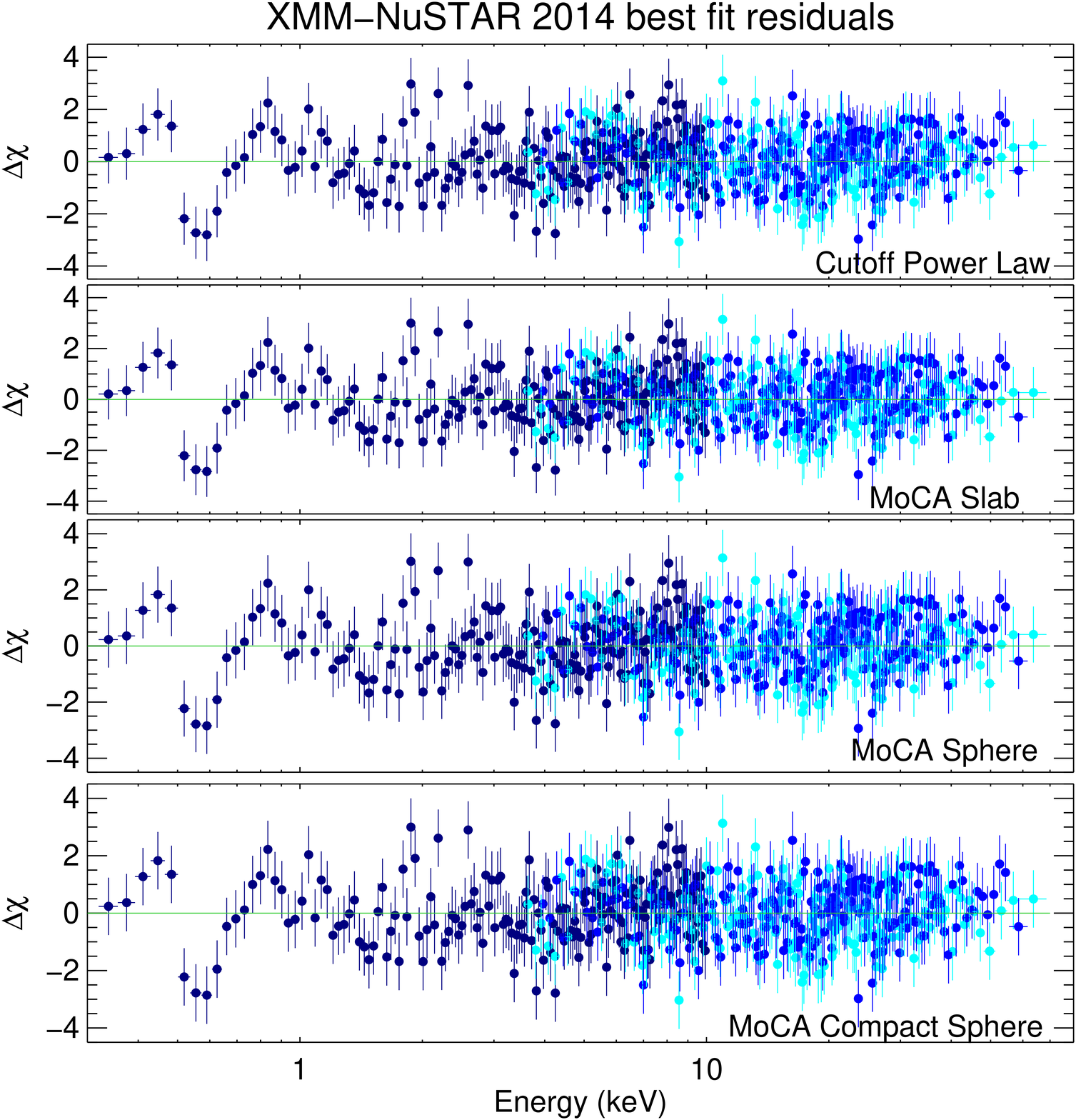, width=0.99\columnwidth}
  \caption{Residuals for the combined 2013 and 2014 XMM+{\it NuSTAR} observations are shown. See text for details about the four models applied. The only residuals present are near to 0.5 keV, which are likely associated to the OI K-shell absorption from our galaxy \citep[see][for further details]{rpb16}.}
  \label{res}
\end{figure*}
\begin{figure*}
 \epsfig{file=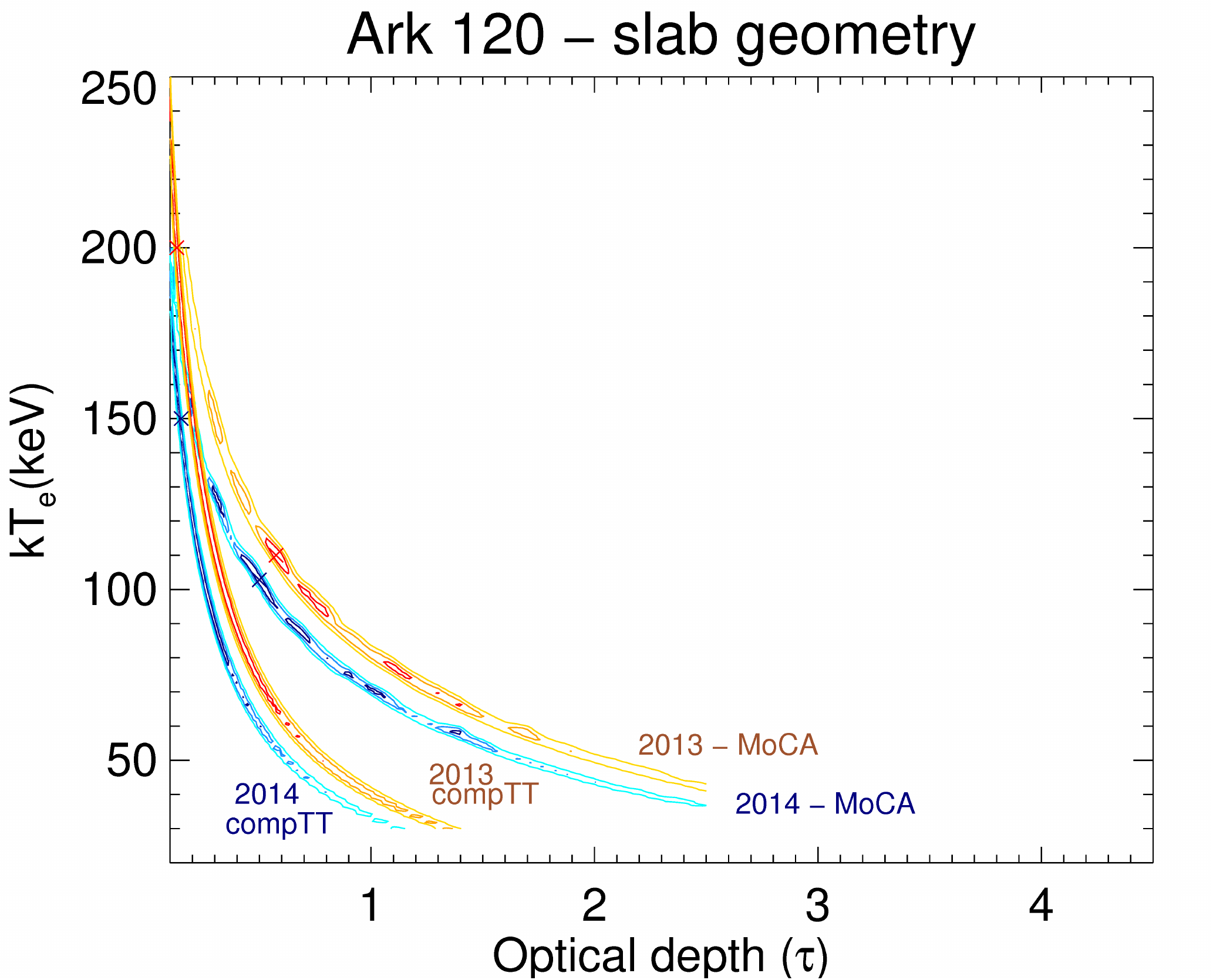, width=\columnwidth}
  \epsfig{file=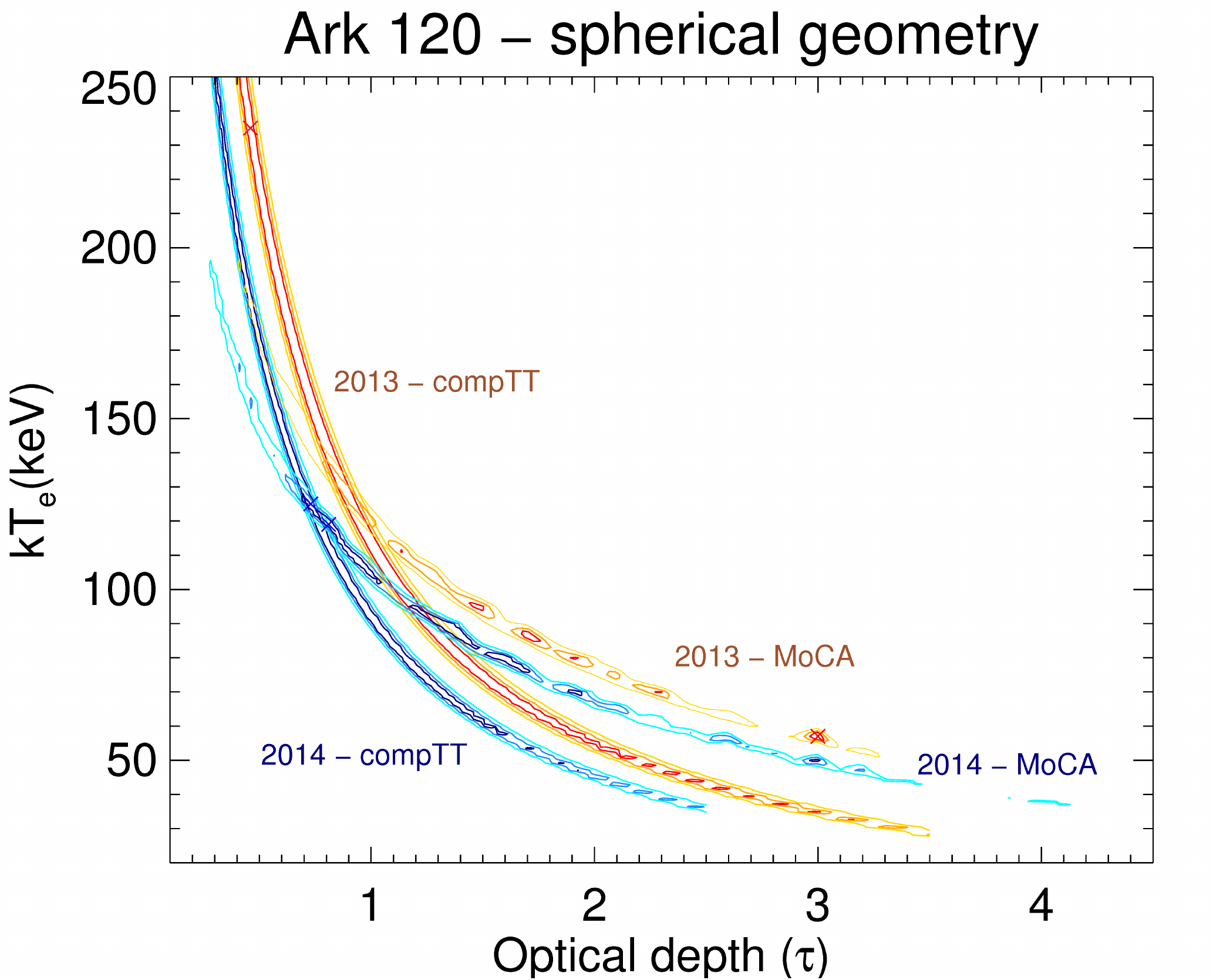, width=\columnwidth}
  \centering
  \epsfig{file=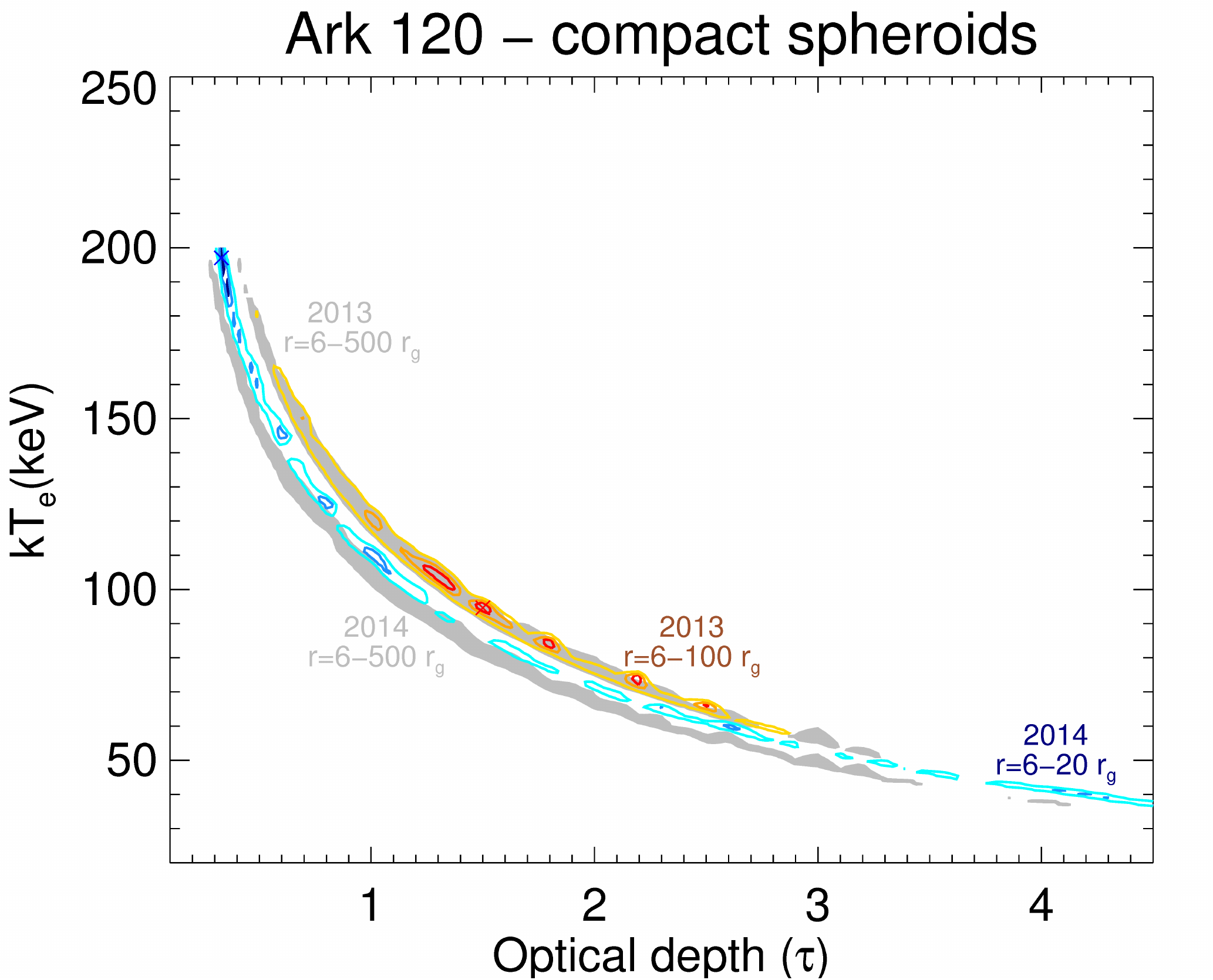, width=\columnwidth}
  \caption{Contour plots between kT$_e$ and $\tau$ for the slab (left panel), spherical (right panel) and compact spherical coronae (bottom panel) are shown, for the 2013 and 2014 best fits. For the 2013 data set yellow, orange and red contours indicate 99\%, 90\% and 68\% confidence levels. For the 2014 data set cyan, light blue and dark blue contours indicate 99\%, 90\% and 68\% confidence levels, respectively. We show, for the sake of visual comparison, the 99\% c.l. contours adopting extended spherical coronae for the 2013 and 2014 observations in gray (bottom panel).}
  \label{slab_cont}
\end{figure*}
When applied to the 2013 and 2014 XMM/{\it NuSTAR} spectra we obtain decent fits, with $\chi^2$/d.o.f.=569/516 and 555/499, respectively. The following parameters are left free to vary: photon index, high energy cutoff and  normalization of the primary power law, kT$_e$, $\tau$ and normalization of the {\sc compTT} component, normalization for the Compton reflection and emission lines. No strong residuals are seen throughout the 0.3-79 keV band (Fig. 2, left and right top panels) and best fit values are in agreement with the ones already discussed in \citet[][b]{mmg14,npr16,preem18}, including cross-calibration constants. For the {\sc compTT} component used to reproduce the soft excess, we find kT$_e=0.39^{+0.08}_{-0.04}$ keV, $\tau=10.1^{+0.7}_{-1.1}$, N=$1.6\pm0.15$ and kT$_e=0.42\pm0.05$ keV, $\tau=9.2^{+0.8}_{-0.6}$, N=$3.6^{+0.3}_{-0.2}$ for the 2013 and 2014 data observations, respectively. \\
\indent
In 2013, Ark 120 was in a flux state F=$2.3\pm0.2\times10^{-11}$ erg cm$^{-2}$ s$^{-1}$, calculated in the 2-10 keV band. Its primary continuum can be well reproduced with a power law with $\Gamma=1.79\pm0.02$ and E$_{\rm c}>200$ keV. On the other hand, on 2014 (March 22) the flux was a factor of $\sim2$ higher, F=$3.9\pm0.1\times10^{-11}$ erg cm$^{-2}$ s$^{-1}$ and the parameters of the nuclear continuum changed to $\Gamma=1.92\pm0.02$ and E$_{\rm c}=300^{+180}_{-100}$ keV: such changes are clearly indicative of a variation of the hot coronal parameters. We therefore substituted the {\sc cutoffpl} power law component with the Comptonisation model {\sc nthcomp} \citep{zjm96,zds99}. We fixed the input seed photon temperature kT$_{bb}$ parameter to the {\sc compTT} one (kT$_{bb}$=15 eV, inferred from the black hole mass and the mean accretion  rate of the source) and left the photon index $\Gamma_{\rm Nth}$, the coronal temperature kT$_{e}$ and the normalization free to vary. We obtain statistically equivalent fits to the ones in which a {\sc cutoffpl} component is used, for both the 2013 and 2014 data sets ($\chi^2$/d.o.f.=574/516 and $\chi^2$/d.o.f.=560/499, respectively): best fit parameters are shown in Table \ref{best_par}. The inferred parameters from the 2013 and 2014 best fits ($\Gamma_{\rm Nth}=1.80\pm0.02$, kT$_{e}>40$ keV and $\Gamma_{\rm Nth}=1.94\pm0.02$, kT$_{e}$=$155^{+350}_{-55}$ keV, respectively) can be translated into optical depths $\tau<2.45$ and $\tau\simeq0.73$, using the formula:
\begin{eqnarray}
\tau = \sqrt{2.25+\frac{3}{((kT_{e}/m_{e}c^2)((\Gamma+0.5)^2-2.25))}}-1.5,
\end{eqnarray}
obtained by inverting Eq. (A1) reported in \citet{zjm96}.

At last, we substituted {\sc NTHcomp} with {\sc compTT}, to directly compare best fit values in slab and spherical geometries with the ones obtained with MoCA. No statistically significant deviations can be found with respect to cutoff power law or {\sc NTHcomp} models; we report best fit values and reduced $\chi^2$ in Table \ref{best_par}, for 2013 and 2014 observations. Contour plots between kT$_e$ and $\tau$ for the two coronal configurations are shown in the top panels of Fig. \ref{slab_cont}.
\subsubsection{\bf Extended coronae}
We then fitted the spectra using the MoCA tables. 
We removed the {\sc cutoffpl} component and replaced it first with the MoCA generated table for the extended slab configuration ({\sc MoCA$_{-}$Slab} model). Variable parameters for the primary continuum are the electron temperature kT$_e$, the optical depth $\tau$ and the model normalization. We obtain statistically equivalent fits for the 2013 and 2014 observations ($\chi^2$/d.o.f.=566/516 and 552/499 respectively) and best fit coronal parameters are reported in Table \ref{best_par}, while residuals are shown in Fig. 2. No statistically significant changes are found for the other parameters. \\
\indent
We then substituted the {\sc MoCA$_{-}$Slab} component with the one generated adopting a geometry in which the scattering electrons are distributed as an extended sphere above the disc ({\sc MoCA$_{-}$Sphere} model). We obtain statistically equivalent fits also in this case ($\chi^2$/d.o.f.=561/516 and 551/499) and best fit parameters are reported in Table \ref{best_par}, while residuals are shown in Fig. 2. Again, no statistically significant changes are found for the other parameters. We conclude, therefore, that none of the two geometries for the hot corona can be preferred from the X-ray spectroscopic analysis.\\
\indent
We show, in Fig. \ref{slab_cont}, the contour plots between kT$_e$ and $\tau$ for the two geometries. Contours from lighter to darker colors indicate 99\%, 90\% and 68\% confidence levels.
It is interesting to note that, for a given geometry, the contour plots for the 2013 and 2014 observations do not overlap. This is not surprising given that a change in the slope of the primary continuum corresponds to changes in the coronal parameters. This is discussed in detail in Middei et al. (in prep.), where the relation between the photon index $\Gamma$ and the Compton parameter obtained from MoCA simulations is presented.\\
\indent
\subsubsection{\bf Compact coronae}
One of the proposed scenarios to explain the variable broad component of the Iron K$\alpha$ in Ark 120 envisages a change in the spatial extension of the hot/warm coronae, which leaves the innermost regions of the accretion disc uncovered. The appearance of the broad, relativistic Iron K$\alpha$ feature from the inner radii of the disc \citep[if fitted with the {\sc relline} model in {\sc xspec} r$^{14c}_{in}=58^{+33}_{-21}$ r$_{g}$ versus r$^{13}_{in}=228^{+72p}_{-130}$ r$_{g}$ are obtained:][]{npr16} could be due to a smaller radius of the hot/warm coronae. This was tested in Porquet et al. 2018b by using the {\sc optxconv} model \citep[based on the {\sc optxagnf} model:][]{ddj12, djm13,jwd12} where we reported a radius of $\sim$73-110 r$_{g}$ in 2013 \citep[where no relativistic component is detected,][]{mmg14} down to $\sim$12-17 r$_{g}$ in 2014.
The model takes into account both the inner warm, thicker corona (to reproduce the soft
excess) and the inner hot, thinner phase (for the hard power law emission), and the emission
form the outer disc. The modeling took into account both disc inclination and relativistic effects.
The variable parameters are mainly the spin of the black hole, the (warm and hot) corona radius 
and the accretion rate.\\
\indent 
We therefore tried to fit the 2013 and 2014 data sets with tables generated adopting a spherical hot corona extended from 6 to 100 r$_g$ and from 6 up to 20 r$_g$, respectively. We find that the reduced $\chi^2$ are comparable with the ones presented in Table \ref{best_par} and we cannot prefer these solutions on statistical ground. It is also worth mentioning that the present version of MoCA does not include GR effects, which may be relevant especially for the most compact corona (see Tamborra 2018b for a more complete discussion on the topic). Contour plots between kT$_e$ and $\tau$ are shown in the bottom panel of Fig. \ref{slab_cont} and while no significant difference is found for the 2013 data set, we can spot a clear discrepancy for the 2014 observation, with respect to extended spherical coronae (see blue contours and gray solid contours for a visual comparison). This effect is further explained in the next section. \\
\begin{figure*}
 \epsfig{file=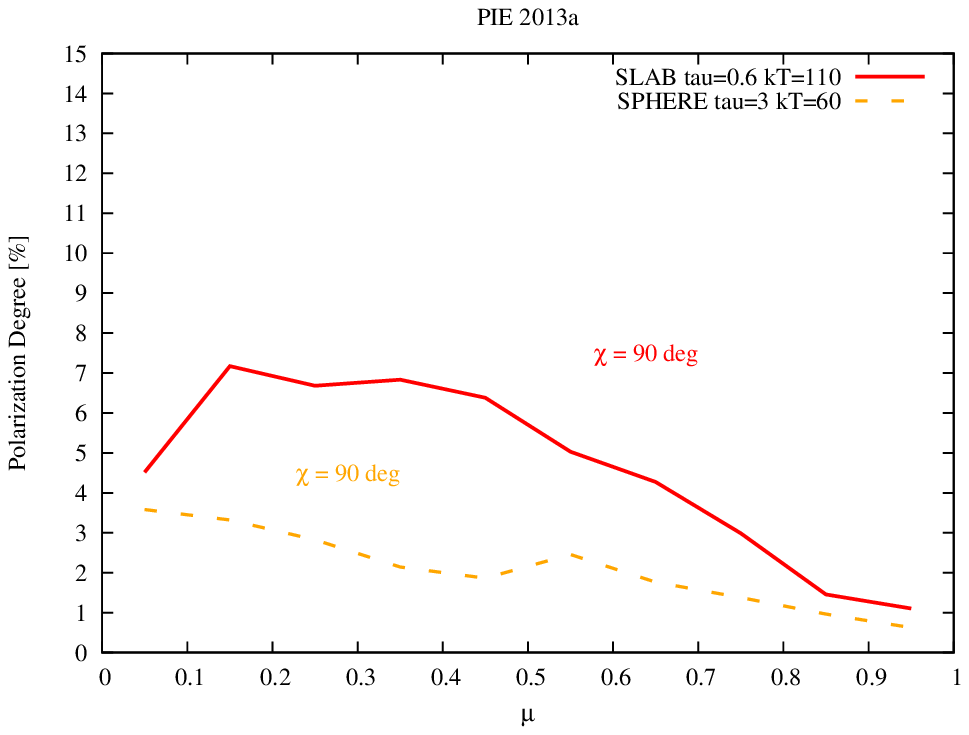, width=\columnwidth}
  \epsfig{file=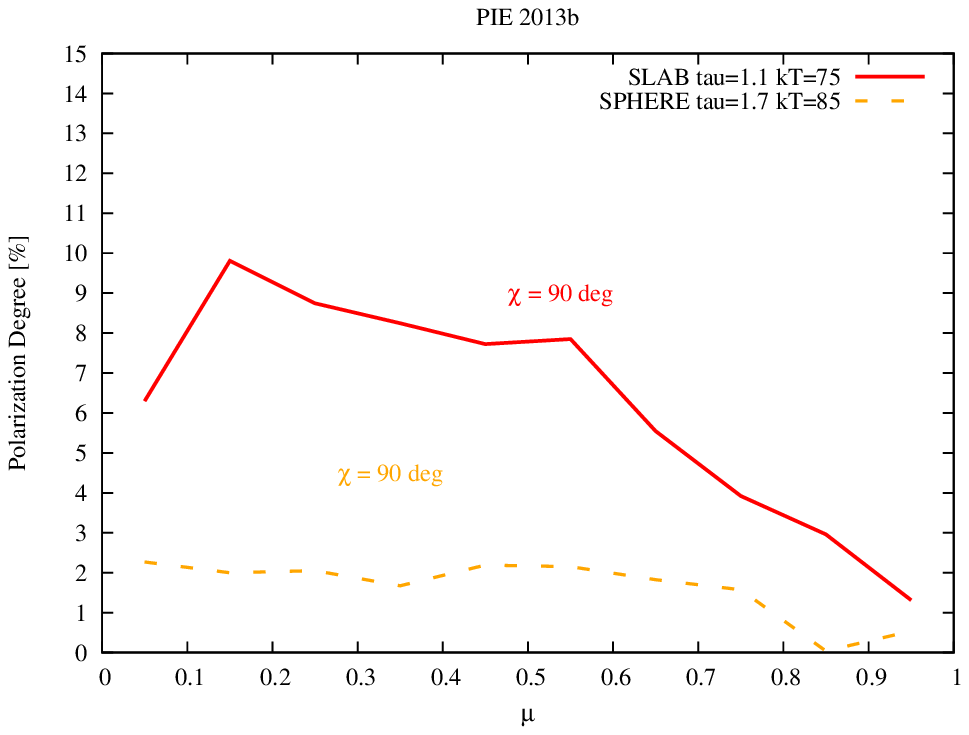, width=\columnwidth}
  \epsfig{file=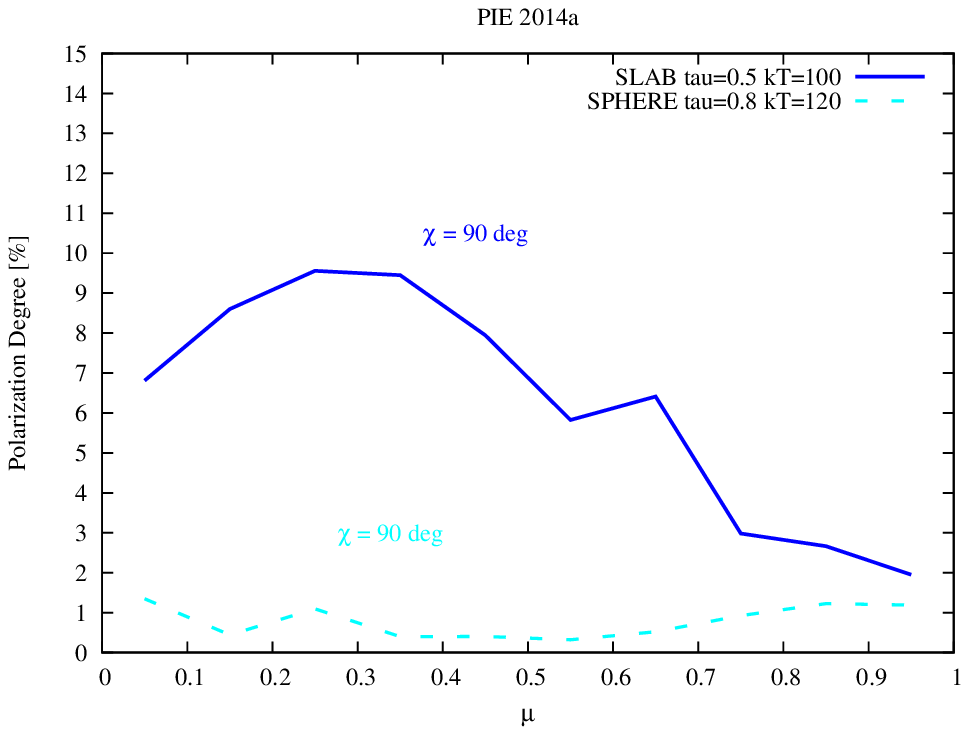, width=\columnwidth}
   \epsfig{file=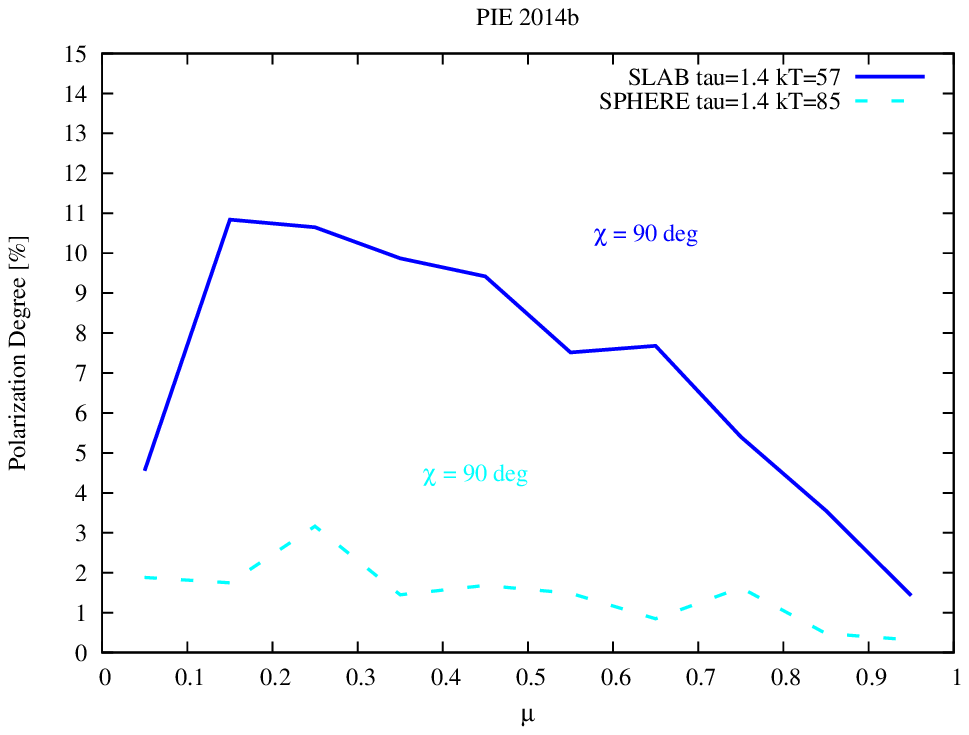, width=\columnwidth}
  \caption{Polarization degree versus the cosine of the inclination angle $\mu$ are shown after integrating the signal in the IXPE operating band (2-8 keV), for the 2013 (top panels) and 2014 (bottom panels) cases and for both extended slab and spherical geometries. Left panels show simulations in which the input kT$_e$ and $\tau$ correspond to best fit values, while we took into account statistically equivalent parameters (i.e. corresponding to relative minima in the fits, see Fig. \ref{slab_cont}) in the right panels.}
  \label{pol_b}
\end{figure*}
\subsection{The geometry of the hot corona}
The different configurations tested in the previous sections provide insights on the geometry of the hot corona in Ark 120. We find a good agreement between the best fit values of the coronal temperature and optical depth obtained with MoCA and the ones inferred using other Comptonisation codes from the literature ({\sc compTT, NTHcomp}).  However, the top panels in Fig. \ref{slab_cont} show that the shape of the contour plots obtained with MoCA are different from the {\sc compTT} ones, indicating that different geometries of the corona lead to different kT$_{e}$-$\tau$ planes. In MoCA, we assumed a slab with a fixed H/R=0.02 and a sphere with radius r=6-500 r$_g$, fully covering the accretion disc. Compared to the {\sc compTT} semi-infinite slab and sphere, a larger value of the input optical depth is needed to recover the same spectral shape, for a fixed electron temperature. \\
The change in normalization between the 2013 and 2014 MoCA best fits ($N^{13}=0.63\pm0.01\times10^{-2}$ and $N^{14}=1.30\pm0.02\times10^{-2}$ ph cm$^{-2}$ s$^{-1}$ keV$^{-1}$ at 1 keV, respectively) can be explained in terms of a corona intercepting more seed photons, resulting in a higher luminosity for the 2014 observation, consistently with the UV fluxes reported in \citet{lpr18} for this object. \\
\indent
We also investigated a physical scenario in which the Comptonisation mechanism occurs in a very compact corona (r=6-20 r$_g$). For the 2014 data set (Fig. \ref{slab_cont}, bottom panel, lower contours), we see that compact coronae need higher optical depths for a given temperature to recover the same spectral shape (or equivalently higher electron temperatures are needed for a given optical depth). This effect is due to the effective number of scatterings that seed photons experience at a given radius. The input spectrum from the disc is assumed to be a multi-temperature black body and photons emitted from the innermost radii will hence be the most energetic ones, despite the geometry of the scattering medium. For values M$_{\rm bh}=1.5\times 10^8$ M$_{\odot}$ and L$_{\rm Bol}$/L$_{\rm Edd}=10\%$, the radial distribution of the seed photons will be therefore peaked at r$\simeq10$ r$_g$, with a negligible contribution from radii above r$\geq100$ r$_g$.  In MoCA, for spherical geometries, the input optical depth is the radial one but, since UV seed photons originate from the accretion disc, photons emitted at different radii will see an effective optical depth $\tau_e$ which will be different from the input optical depth $\tau$.  The effective optical depth of seed photons from the innermost regions decreases significantly in 6-20 r$_g$ coronae. From our calculations, in a kT=100 keV and $\tau$=1 configuration, while photons emitted below 10 r$_g$ experience on average N$_s\simeq$0.75 scatterings in the 6-500 r$_g$ and 6-100 r$_g$ cases, this number drops to N$_s\simeq$0.62 for 6-20 r$_g$ coronae. This effect is further confirmed by comparing our simulated spectra for extended and compact coronae: we observe steeper continua (and hence lower $\tau_e$) for 6-20 r$_g$ coronae. On average, the discrepancy between the photon indices is $\Delta\Gamma\simeq0.1$, as discussed in \citet[][]{tmb18}. Therefore we can conclude that thicker coronae are needed for a given temperature (or vice versa higher temperatures for a given $\tau$) to recover the same spectral shape, in the case of 6-20 r$_g$ spherical coronae. 
\section{\bf X-ray polarization}
Compton scattering, as any scattering mechanism produces a linear polarization, which is strongly sensitive to the symmetry of the scattering medium (i.e. the hot corona) and of the radiation field. For the geometries considered here, we expect a polarization which is perpendicular to the plane of the disc (from now-on we will refer to it as vertical) with the slab showing a higher degree of polarization with respect to the sphere. The thermal emission from the disc is assumed to be polarized parallel to the disc plane (we will refer to it as horizontal) according to Chandrasekhar results for a pure electron scattering, optically thick atmospheres in plane-parallel approximation; the polarization degrees goes from zero for a face-on view of the disc, to almost 12\% for an edge-on view \citep{chandra60}.\\
\indent
We therefore chose the best fit coronal parameters for the 2013 and 2014 XMM-{\it NuSTAR} observations of Ark 120 and calculated the expected X-ray polarization as function of the inclination angle, integrated between 2 and 8 keV.\\
\indent
For the 2013 observations, we selected kT$_e$=110 keV-$\tau=0.6$ for the extended slab and kT$_e$=60 keV-$\tau=3.0$ for the extended spherical geometry (Fig. \ref{pol_b}, top-left panel). These values correspond to the absolute minima of the fits (see Table \ref{best_par}). We also chose two more pairs of parameters, corresponding to the relative minima of the fits (see contour plots presented in Fig. \ref{slab_cont}). These additional pairs are kT$_e$=75 keV-$\tau=1.1$ and kT$_e$=85 keV-$\tau=1.7$ for slab spherical geometries, respectively (Fig. \ref{pol_b}, top-right panel).\\
\indent
We followed the same approach for the 2014 observations and selected kT$_e$=100 keV-$\tau=0.5$ for the extended slab and kT$_e$=120 keV-$\tau=0.8$ for the extended spherical geometry as absolute minima of the fits (Fig. \ref{pol_b}, bottom-left panel). As secondary sets of parameters we simulated data using kT$_e$=57 keV-$\tau=1.4$ and  kT$_e$=85 keV-$\tau=1.4$ for the slab and the sphere, respectively (Fig. \ref{pol_b}, bottom-right panel).\\ 
\subsection{Polarization angle}
The polarization angle is always consistent with $\chi=\pi/2$ for every considered configuration.\\
Compton scattering produces a vertical polarization, i.e. $\chi=\pi/2$, in both geometries, with the angle measured with respect to the spin axis.
Even though some photons are initially polarized horizontally, the flip in the polarization angle \citep[seen in][for instance]{sk10} is not visible in the IXPE band where only photons which experienced multiple scattering are present and their polarization is already vertical.
\subsection{Polarization degree}
We show, in Fig. \ref{pol_b}, the expected polarization degree versus the cosine of the inclination angle ($\mu=\cos\theta$, with $\theta$ defined from the spin axis) in the four considered cases, assuming radiation integrated between 2-8 keV  (corresponding to the IXPE operating energy band). Simulations for the 2013 and 2014 data sets are shown in the top and bottom panels, respectively. The expected signal arising from the slab configuration is plotted using a solid line while dashed lines are used for the spherical configurations. The polarization for the spherical corona is always lower than for the slab, in every simulated scenario, due to obvious symmetry reasons. For a given geometry we may also notice that thicker coronae will produce a larger polarization signal: the more the UV photons are Compton scattered, the more they will be polarized. \\
The maximum of the polarization degree is obtained towards high inclination angles (i.e. low values of $\mu$). On the other hand, when the system is observed face-on, for $\mu$=1, it is perfectly symmetrical for both geometries and the polarization degree approaches zero in all scenarios, as expected. It is important to note that the polarization degree is not a monothonic function of $\mu$, but has a maximum around $\mu\simeq0.2$, as already found by \citet[][Fig. 2]{ang69} for a $\tau=2$ scattering spheroid.

\subsection{\bf Polarization measurements with IXPE}
To calculate the exposure time needed to constrain the coronal geometry with IXPE, we assumed a model composed of a power law with photon index $\Gamma=1.94$ and a 2-8 keV flux of $3.3\times10^{-11}$ erg cm$^{-2}$ s$^{-1}$. From Fig. \ref{pol_b} we can clearly see that the lowest values for the polarization degree are obtained in spherical coronae, ranging from 1\% up to 4\%. Much larger values are found in the slab geometry. The minimum polarization degree measurable at a certain confidence level \citep[MDP, see][for details]{eow12} is given by:
\begin{equation}
 MDP=\sqrt{-2 \ln(1-CL)}\sqrt{2}\frac{\sqrt{C_S+C_B}}{C_S<\mu>}
\end{equation}
where $C_S$ are source counts, $C_B$ background counts, $CL$ is the confidence level, and $<\mu>$ is the source-count-weighted modulation factor. In order to achieve an MDP of 2.5$\%$ at 99\% c.l. in an IXPE observation of Ark 120, an exposure time t${}_{\rm exp}\sim$800 ks is needed\footnote{The IXPE sensitivity calculator can be found at \href{https://ixpe.msfc.nasa.gov/for_scientists/pimms/}{https://ixpe.msfc.nasa.gov/for$_{-}$scientists/pimms/}}. 
The source is a bare Seyfert galaxy and its negligible intrinsic X-ray absorption \citep[N$_{\rm H}<3.4\times10^{19}$ cm$^{-2}$:][]{rpb16} suggests a low inclination for the accretion disc-corona system. In fact, an inclination angle $i$=26$^{\circ}$ is measured for the host galaxy \citep{nhc95} but a misalignment between the X-ray obscuring material and the accretion disc cannot be excluded a priori \citep{le10}. Different inclination angles of the accretion disc have been inferred via Iron K$\alpha$ line spectroscopy \citep[$i$=35$\pm2$ and $i$=57$^{+5}_{-12}$ degrees:][respectively]{preem18,nfr11} or using the {\sc optxconv} model which accounts for the warm and hot Comptonisation ($i$=20-35 degrees, Porquet et al., 2018). These ranges in inclination angles translate into a global range of $\mu$=0.45-0.9 \citep[see][for a complete list of measurements]{fma16}, which clearly allows us to measure a polarization signal and to distinguish between a slab or spherical geometry (see Fig. \ref{pol_b}).
\section{\bf Conclusions}
We presented the first application of the {\sc MoCA} Comptonisation code to real X-ray data to model the hot corona for the bare Seyfert galaxy Ark 120. Our results can be summarized as follows:
\begin{itemize}
 \item the 2013 and 2014 simultaneous XMM and {\it NuSTAR} observations of Ark 120 have been analyzed and different models have been tested, to reproduce the X-ray shape of its nuclear continuum in order to characterize the hot corona. No statistically significant difference is found between a phenomenological model (cutoff power law), {\sc NTHcomp}, {\sc compTT} and {\sc MoCA};\\
 \item {\sc xspec} readable tables based on {\sc MoCA} have been generated for multiple geometries of the hot corona. We considered an extended slab above the accretion disc (r=6-500 r$_g$), an extended sphere (r=6-500 r$_g$) and two compact spheres (r=6-20 r$_g$ and r=6-100 r$_g$). All the tested geometries for the hot coronae are statistically equivalent, based on spectroscopy solely, for both the 2013 and 2014 data sets. We found that thicker hot coronae are needed for a given temperature (or equivalently higher temperatures for a given $\tau$) to recover the same spectral shape, in the case of 6-20 r$_g$ spherical coronae;\\
 \item a general agreement is found between the best fit values of the hot coronal parameters obtained with MoCA and the ones inferred using other Comptonisation codes from the literature ({\sc compTT, NTHcomp}). However, the comparison between the best fit kT$_e$-$\tau$ contour plots stress the importance of MoCA in modeling the geometry of the scattering material;\\
 \item best fit values from the spectral fits have been used to run simulations to calculate the polarization properties, when integrated in the IXPE energy band. A difference between the two geometries for the hot corona clearly emerges, due to the higher asymmetry of the extended slab compared to the extended sphere configuration;\\
 \item  we estimated that the two geometries for the hot corona can be distinguished with a 
t${}_{\rm exp}\sim$800 ks with IXPE, corresponding to a MDP$_{99}$=2.5$\%$.
 
\end{itemize}

\section*{ACKNOWLEDGEMENTS}
{We thank the anonymous referee for her/his comments.} AM thanks Craig Gordon for useful suggestions about generating the MoCA {\sc xspec} tables.
  AM, GM acknowledge financial support from the Italian Space Agency under grant ASI/INAF I/037/12/0-011/13. SB acknowledges financial support from the Italian Space Agency under grant ASI-INAF I/037/12/0. AM, GM, SB, RM, AT acknowledge financial support from ASI-INAF 2017-12-H.0. AL acknowledges support from STFC consolidated grant ST/M001040/1. FM would like to thank the Centre national d'\`etudes spatiales (CNES) who funded this project through to the post-doctoral grant ‘Probing the geometry and physics of active galactic nuclei with ultraviolet and X-ray polarized radiative transfer’. EN acknowledges funding from the European Union's Horizon 2020 research and innovation programme under the Marie Sk\l{l}odowska-Curie grant agreement No. 664931. 

\bibliographystyle{aa}
\bibliographystyle{aa}
\bibliography{sbs} 

\begin{thebibliography}{67}
\expandafter\ifx\csname natexlab\endcsname\relax\def\natexlab#1{#1}\fi

\bibitem[{{Angel}(1969)}]{ang69}
{Angel}, J.~R.~P. 1969, \apj, 158, 219

\bibitem[{{Arnaud}(1996)}]{xspec}
{Arnaud}, K.~A. 1996, in ASP Conf. Ser. 101: Astronomical Data Analysis
  Software and Systems V, 17

\bibitem[{{Arnaud} {et~al.}(1985){Arnaud}, {Branduardi-Raymont}, {Culhane},
  {Fabian}, {Hazard}, {McGlynn}, {Shafer}, {Tennant}, \& {Ward}}]{arn85}
{Arnaud}, K.~A., {Branduardi-Raymont}, G., {Culhane}, J.~L., {et~al.} 1985,
  \mnras, 217, 105

\bibitem[{{Beheshtipour} {et~al.}(2017){Beheshtipour}, {Krawczynski}, \&
  {Malzac}}]{bkm17}
{Beheshtipour}, B., {Krawczynski}, H., \& {Malzac}, J. 2017, \apj, 850, 14

\bibitem[{{Beloborodov}(1999)}]{bel99}
{Beloborodov}, A.~M. 1999, in ASP Conf. Ser. 161: High Energy Processes in
  Accreting Black Holes, 295--+

\bibitem[{{Bianchi} {et~al.}(2009){Bianchi}, {Guainazzi}, {Matt}, {Fonseca
  Bonilla}, \& {Ponti}}]{bianchi09}
{Bianchi}, S., {Guainazzi}, M., {Matt}, G., {Fonseca Bonilla}, N., \& {Ponti},
  G. 2009, \aap, 495, 421

\bibitem[{{Chandrasekhar}(1960)}]{chandra60}
{Chandrasekhar}, S. 1960, {Radiative transfer} (New York: Dover, 1960)

\bibitem[{{Crenshaw} {et~al.}(1999){Crenshaw}, {Kraemer}, {Boggess}, {Maran},
  {Mushotzky}, \& {Wu}}]{ckb99}
{Crenshaw}, D.~M., {Kraemer}, S.~B., {Boggess}, A., {et~al.} 1999, \apj, 516,
  750

\bibitem[{{Dauser} {et~al.}(2014){Dauser}, {Garc{\'{\i}}a}, {Parker}, {Fabian},
  \& {Wilms}}]{dgp14}
{Dauser}, T., {Garc{\'{\i}}a}, J., {Parker}, M.~L., {Fabian}, A.~C., \&
  {Wilms}, J. 2014, \mnras, 444, L100

\bibitem[{{De Marco} {et~al.}(2013){De Marco}, {Ponti}, {Cappi}, {Dadina},
  {Uttley}, {Cackett}, {Fabian}, \& {Miniutti}}]{demarco13}
{De Marco}, B., {Ponti}, G., {Cappi}, M., {et~al.} 2013, \mnras, 431, 2441

\bibitem[{{Done} {et~al.}(2012){Done}, {Davis}, {Jin}, {Blaes}, \&
  {Ward}}]{ddj12}
{Done}, C., {Davis}, S.~W., {Jin}, C., {Blaes}, O., \& {Ward}, M. 2012, \mnras,
  420, 1848

\bibitem[{{Done} {et~al.}(2013){Done}, {Jin}, {Middleton}, \& {Ward}}]{djm13}
{Done}, C., {Jin}, C., {Middleton}, M., \& {Ward}, M. 2013, \mnras, 434, 1955

\bibitem[{{Elsner} {et~al.}(2012){Elsner}, {O'Dell}, \& {Weisskopf}}]{eow12}
{Elsner}, R.~F., {O'Dell}, S.~L., \& {Weisskopf}, M.~C. 2012, in \procspie,
  Vol. 8443, Space Telescopes and Instrumentation 2012: Ultraviolet to Gamma
  Ray, 84434N

\bibitem[{{Fabian} {et~al.}(2017){Fabian}, {Lohfink}, {Belmont}, {Malzac}, \&
  {Coppi}}]{flb17}
{Fabian}, A.~C., {Lohfink}, A., {Belmont}, R., {Malzac}, J., \& {Coppi}, P.
  2017, \mnras, 467, 2566

\bibitem[{{Fabian} {et~al.}(2015){Fabian}, {Lohfink}, {Kara}, {Parker},
  {Vasudevan}, \& {Reynolds}}]{flk15}
{Fabian}, A.~C., {Lohfink}, A., {Kara}, E., {et~al.} 2015, \mnras, 451, 4375

\bibitem[{{Gabriel} {et~al.}(2004){Gabriel}, {Denby}, {Fyfe}, {Hoar}, {Ibarra},
  {Ojero}, {Osborne}, {Saxton}, {Lammers}, \& {Vacanti}}]{gabr04}
{Gabriel}, C., {Denby}, M., {Fyfe}, D.~J., {et~al.} 2004, in Astronomical
  Society of the Pacific Conference Series, Vol. 314, Astronomical Data
  Analysis Software and Systems (ADASS) XIII, ed. {F.~Ochsenbein, M.~G.~Allen,
  \& D.~Egret}, 759--+

\bibitem[{{Garc{\'{\i}}a} {et~al.}(2014){Garc{\'{\i}}a}, {Dauser}, {Lohfink},
  {Kallman}, {Steiner}, {McClintock}, {Brenneman}, {Wilms}, {Eikmann},
  {Reynolds}, \& {Tombesi}}]{gdl14}
{Garc{\'{\i}}a}, J., {Dauser}, T., {Lohfink}, A., {et~al.} 2014, \apj, 782, 76

\bibitem[{{Haardt} \& {Maraschi}(1991)}]{hm91}
{Haardt}, F. \& {Maraschi}, L. 1991, \apjl, 380, L51

\bibitem[{{Haardt} {et~al.}(1994){Haardt}, {Maraschi}, \& {Ghisellini}}]{hmg94}
{Haardt}, F., {Maraschi}, L., \& {Ghisellini}, G. 1994, \apjl, 432, L95

\bibitem[{{Harrison} {et~al.}(2013){Harrison}, {Craig}, {Christensen},
  {Hailey}, {Zhang}, {Boggs}, {Stern}, {Cook}, {Forster}, {Giommi},
  {Grefenstette}, {Kim}, {Kitaguchi}, {Koglin}, {Madsen}, {Mao}, {Miyasaka},
  {Mori}, {Perri}, {Pivovaroff}, {Puccetti}, {Rana}, {Westergaard}, {Willis},
  {Zoglauer}, {An}, {Bachetti}, {Barri{\`e}re}, {Bellm}, {Bhalerao},
  {Brejnholt}, {Fuerst}, {Liebe}, {Markwardt}, {Nynka}, {Vogel}, {Walton},
  {Wik}, {Alexander}, {Cominsky}, {Hornschemeier}, {Hornstrup}, {Kaspi},
  {Madejski}, {Matt}, {Molendi}, {Smith}, {Tomsick}, {Ajello}, {Ballantyne},
  {Balokovi{\'c}}, {Barret}, {Bauer}, {Blandford}, {Niel Brandt}, {Brenneman},
  {Chiang}, {Chakrabarty}, {Chenevez}, {Comastri}, {Dufour}, {Elvis}, {Fabian},
  {Farrah}, {Fryer}, {Gotthelf}, {Grindlay}, {Helfand}, {Krivonos}, {Meier},
  {Miller}, {Natalucci}, {Ogle}, {Ofek}, {Ptak}, {Reynolds}, {Rigby},
  {Tagliaferri}, {Thorsett}, {Treister}, \& {Urry}}]{nustar}
{Harrison}, F.~A., {Craig}, W.~W., {Christensen}, F.~E., {et~al.} 2013, \apj,
  770, 103

\bibitem[{{Jansen} {et~al.}(2001){Jansen}, {Lumb}, {Altieri}, {Clavel}, {Ehle},
  {Erd}, {Gabriel}, {Guainazzi}, {Gondoin}, {Much}, {Munoz}, {Santos},
  {Schartel}, {Texier}, \& {Vacanti}}]{xmm}
{Jansen}, F., {Lumb}, D., {Altieri}, B., {et~al.} 2001, \aap, 365, L1

\bibitem[{{Jin} {et~al.}(2012){Jin}, {Ward}, {Done}, \& {Gelbord}}]{jwd12}
{Jin}, C., {Ward}, M., {Done}, C., \& {Gelbord}, J. 2012, \mnras, 420, 1825

\bibitem[{{Kalberla} {et~al.}(2005){Kalberla}, {Burton}, {Hartmann}, {Arnal},
  {Bajaja}, {Morras}, \& {P{\"o}ppel}}]{kalberla05}
{Kalberla}, P.~M.~W., {Burton}, W.~B., {Hartmann}, D., {et~al.} 2005, \aap,
  440, 775

\bibitem[{{Kara} {et~al.}(2016){Kara}, {Alston}, {Fabian}, {Cackett}, {Uttley},
  {Reynolds}, \& {Zoghbi}}]{kaf16}
{Kara}, E., {Alston}, W.~N., {Fabian}, A.~C., {et~al.} 2016, \mnras, 462, 511

\bibitem[{{Lawrence} \& {Elvis}(2010)}]{le10}
{Lawrence}, A. \& {Elvis}, M. 2010, \apj, 714, 561

\bibitem[{{Lightman} \& {Zdziarski}(1987)}]{lz87}
{Lightman}, A.~P. \& {Zdziarski}, A.~A. 1987, \apj, 319, 643

\bibitem[{{Lobban} {et~al.}(2018){Lobban}, {Porquet}, {Reeves}, {Markowitz},
  {Nardini}, \& {Grosso}}]{lpr18}
{Lobban}, A.~P., {Porquet}, D., {Reeves}, J.~N., {et~al.} 2018, \mnras, 474,
  3237

\bibitem[{{Magdziarz} {et~al.}(1998){Magdziarz}, {Blaes}, {Zdziarski},
  {Johnson}, \& {Smith}}]{mbz98}
{Magdziarz}, P., {Blaes}, O.~M., {Zdziarski}, A.~A., {Johnson}, W.~N., \&
  {Smith}, D.~A. 1998, \mnras, 301, 179

\bibitem[{{Marin}(2016)}]{fma16}
{Marin}, F. 2016, \mnras, 460, 3679

\bibitem[{{Matt} {et~al.}(2014){Matt}, {Marinucci}, {Guainazzi}, {Brenneman},
  {Elvis}, {Lohfink}, {Ar{\`e}valo}, {Boggs}, {Cappi}, {Christensen}, {Craig},
  \& {Fabian}}]{mmg14}
{Matt}, G., {Marinucci}, A., {Guainazzi}, M., {et~al.} 2014, \mnras, 439, 3016

\bibitem[{{Nardini} {et~al.}(2011){Nardini}, {Fabian}, {Reis}, \&
  {Walton}}]{nfr11}
{Nardini}, E., {Fabian}, A.~C., {Reis}, R.~C., \& {Walton}, D.~J. 2011, \mnras,
  410, 1251

\bibitem[{{Nardini} {et~al.}(2016){Nardini}, {Porquet}, {Reeves}, {Braito},
  {Lobban}, \& {Matt}}]{npr16}
{Nardini}, E., {Porquet}, D., {Reeves}, J.~N., {et~al.} 2016, \apj, 832, 45

\bibitem[{{Nordgren} {et~al.}(1995){Nordgren}, {Helou}, {Chengalur}, {Terzian},
  \& {Khachikian}}]{nhc95}
{Nordgren}, T.~E., {Helou}, G., {Chengalur}, J.~N., {Terzian}, Y., \&
  {Khachikian}, E. 1995, \apjs, 99, 461

\bibitem[{{Osterbrock} \& {Phillips}(1977)}]{op77}
{Osterbrock}, D.~E. \& {Phillips}, M.~M. 1977, \pasp, 89, 251

\bibitem[{{Peterson} {et~al.}(2004){Peterson}, {Ferrarese}, {Gilbert}, {Kaspi},
  {Malkan}, {Maoz}, {Merritt}, {Netzer}, {Onken}, {Pogge}, {Vestergaard}, \&
  {Wandel}}]{pet04}
{Peterson}, B.~M., {Ferrarese}, L., {Gilbert}, K.~M., {et~al.} 2004, \apj, 613,
  682

\bibitem[{{Petrucci} {et~al.}(2001){Petrucci}, {Haardt}, {Maraschi}, {Grandi},
  {Malzac}, {Matt}, {Nicastro}, {Piro}, {Perola}, \& {De Rosa}}]{phm01}
{Petrucci}, P.~O., {Haardt}, F., {Maraschi}, L., {et~al.} 2001, \apj, 556, 716

\bibitem[{{Petrucci} {et~al.}(2000){Petrucci}, {Haardt}, {Maraschi}, {Grandi},
  {Matt}, {Nicastro}, {Piro}, {Perola}, \& {De Rosa}}]{petr00}
{Petrucci}, P.~O., {Haardt}, F., {Maraschi}, L., {et~al.} 2000, \apj, 540, 131

\bibitem[{{Petrucci} {et~al.}(2013){Petrucci}, {Paltani}, {Malzac}, {Kaastra},
  {Cappi}, {Ponti}, {De Marco}, {Kriss}, {Steenbrugge}, {Bianchi},
  {Branduardi-Raymont}, {Mehdipour}, {Costantini}, {Dadina}, \&
  {Lubi{\'n}ski}}]{ppm13}
{Petrucci}, P.-O., {Paltani}, S., {Malzac}, J., {et~al.} 2013, \aap, 549, A73

\bibitem[{{Petrucci} {et~al.}(2018){Petrucci}, {Ursini}, {De Rosa}, {Bianchi},
  {Cappi}, {Matt}, {Dadina}, \& {Malzac}}]{pud18}
{Petrucci}, P.-O., {Ursini}, F., {De Rosa}, A., {et~al.} 2018, \aap, 611, A59

\bibitem[{{Piconcelli} {et~al.}(2004){Piconcelli}, {Jimenez-Bail{\' o}n},
  {Guainazzi}, {Schartel}, {Rodr{\'{\i}}guez-Pascual}, \& {Santos-Lle{\'
  o}}}]{pico04}
{Piconcelli}, E., {Jimenez-Bail{\' o}n}, E., {Guainazzi}, M., {et~al.} 2004,
  \mnras, 351, 161

\bibitem[{{Porquet} {et~al.}(2018a){Porquet}, {Reeves}, {Matt}, {Marinucci},
  {Nardini}, {Braito}, {Lobban}, {Ballantyne}, {Boggs}, {Christensen},
  {Dauser}, {Farrah}, {Garcia}, {Hailey}, {Harrison}, {Stern}, {Tortosa},
  {Ursini}, \& {Zhang}}]{preem18}
{Porquet}, D., {Reeves}, J.~N., {Matt}, G., {et~al.} 2018, \aap, 609, A42

\bibitem[{{Porquet} {et~al.}(2004){Porquet}, {Reeves}, {O'Brien}, \&
  {Brinkmann}}]{pro04}
{Porquet}, D., {Reeves}, J.~N., {O'Brien}, P., \& {Brinkmann}, W. 2004, \aap,
  422, 85

\bibitem[{{Poutanen} \& {Svensson}(1996)}]{ps96}
{Poutanen}, J. \& {Svensson}, R. 1996, \apj, 470, 249

\bibitem[{{Reeves} {et~al.}(2016){Reeves}, {Porquet}, {Braito}, {Nardini},
  {Lobban}, \& {Turner}}]{rpb16}
{Reeves}, J.~N., {Porquet}, D., {Braito}, V., {et~al.} 2016, \apj, 828, 98

\bibitem[{{Reis} \& {Miller}(2013)}]{rm13}
{Reis}, R.~C. \& {Miller}, J.~M. 2013, \apjl, 769, L7

\bibitem[{{Risaliti} {et~al.}(2011){Risaliti}, {Nardini}, {Elvis}, {Brenneman},
  \& {Salvati}}]{rne11}
{Risaliti}, G., {Nardini}, E., {Elvis}, M., {Brenneman}, L., \& {Salvati}, M.
  2011, \mnras, 417, 178

\bibitem[{{Risaliti} {et~al.}(2009){Risaliti}, {Young}, \& {Elvis}}]{rye09}
{Risaliti}, G., {Young}, M., \& {Elvis}, M. 2009, \apjl, 700, L6

\bibitem[{{R{\'o}{\.z}a{\'n}ska} {et~al.}(2015){R{\'o}{\.z}a{\'n}ska},
  {Malzac}, {Belmont}, {Czerny}, \& {Petrucci}}]{rmb15}
{R{\'o}{\.z}a{\'n}ska}, A., {Malzac}, J., {Belmont}, R., {Czerny}, B., \&
  {Petrucci}, P.-O. 2015, \aap, 580, A77

\bibitem[{{Rybicki} \& {Lightman}(1979)}]{rl79}
{Rybicki}, G.~B. \& {Lightman}, A.~P. 1979, {Radiative processes in
  astrophysics}

\bibitem[{{Sanfrutos} {et~al.}(2013){Sanfrutos}, {Miniutti},
  {Ag{\'{\i}}s-Gonz{\'a}lez}, {Fabian}, {Miller}, {Panessa}, \&
  {Zoghbi}}]{sanmi13}
{Sanfrutos}, M., {Miniutti}, G., {Ag{\'{\i}}s-Gonz{\'a}lez}, B., {et~al.} 2013,
  ArXiv e-prints [\eprint[arXiv]{1309.1092}]

\bibitem[{{Schnittman} \& {Krolik}(2010)}]{sk10}
{Schnittman}, J.~D. \& {Krolik}, J.~H. 2010, \apj, 712, 908

\bibitem[{{Shapiro} {et~al.}(1976){Shapiro}, {Lightman}, \& {Eardley}}]{sle76}
{Shapiro}, S.~L., {Lightman}, A.~P., \& {Eardley}, D.~M. 1976, \apj, 204, 187

\bibitem[{{Shemmer} {et~al.}(2006){Shemmer}, {Brandt}, {Netzer}, {Maiolino}, \&
  {Kaspi}}]{shem06}
{Shemmer}, O., {Brandt}, W.~N., {Netzer}, H., {Maiolino}, R., \& {Kaspi}, S.
  2006, \apjl, 646, L29

\bibitem[{{Singh} {et~al.}(1985){Singh}, {Garmire}, \& {Nousek}}]{sgn85}
{Singh}, K.~P., {Garmire}, G.~P., \& {Nousek}, J. 1985, \apj, 297, 633

\bibitem[{{Sobolewska} \& {Papadakis}(2009)}]{sp09}
{Sobolewska}, M.~A. \& {Papadakis}, I.~E. 2009, \mnras, 399, 1597

\bibitem[{{Str{\"u}der} {et~al.}(2001){Str{\"u}der}, {Briel}, {Dennerl},
  {Hartmann}, {Kendziorra}, {Meidinger}, {Pfeffermann}, {Reppin}, {Aschenbach},
  {Bornemann}, {Br{\" a}uninger}, {Burkert}, \& {Elender}}]{struder01}
{Str{\"u}der}, L., {Briel}, U., {Dennerl}, K., {et~al.} 2001, \aap, 365, L18

\bibitem[{{Sunyaev} \& {Titarchuk}(1980)}]{st80}
{Sunyaev}, R.~A. \& {Titarchuk}, L.~G. 1980, \aap, 86, 121

\bibitem[{{Tamborra} {et~al.}(2018){Tamborra}, {Matt}, {Bianchi}, \& {Dov{\v
  c}iak}}]{tmb18}
{Tamborra}, F., {Matt}, G., {Bianchi}, S., \& {Dov{\v c}iak}, M. 2018, ArXiv
  e-prints [\eprint[arXiv]{1808.07399}]

\bibitem[{{Titarchuk}(1994)}]{tit94}
{Titarchuk}, L. 1994, \apj, 434, 570

\bibitem[{{Tortosa} {et~al.}(2018){Tortosa}, {Bianchi}, {Marinucci}, {Matt}, \&
  {Petrucci}}]{tbm18}
{Tortosa}, A., {Bianchi}, S., {Marinucci}, A., {Matt}, G., \& {Petrucci}, P.~O.
  2018, \aap, 614, A37

\bibitem[{{Turner} {et~al.}(2001){Turner}, {Abbey}, {Arnaud}, {Balasini},
  {Barbera}, {Belsole}, {Bennie}, {Bernard}, {Bignami}, {Boer}, {Briel},
  {Butler}, {Cara}, {Chabaud}, {Cole}, {Collura}, {Conte}, {Cros}, \&
  {Denby}}]{turner01}
{Turner}, M.~J.~L., {Abbey}, A., {Arnaud}, M., {et~al.} 2001, \aap, 365, L27

\bibitem[{{Ursini} {et~al.}(2016){Ursini}, {Petrucci}, {Matt}, {Bianchi},
  {Cappi}, {De Marco}, {De Rosa}, {Malzac}, {Marinucci}, {Ponti}, \&
  {Tortosa}}]{upm16}
{Ursini}, F., {Petrucci}, P.-O., {Matt}, G., {et~al.} 2016, \mnras, 463, 382

\bibitem[{{Vaughan} {et~al.}(2004){Vaughan}, {Fabian}, {Ballantyne}, {De Rosa},
  {Piro}, \& {Matt}}]{vfb04}
{Vaughan}, S., {Fabian}, A.~C., {Ballantyne}, D.~R., {et~al.} 2004, \mnras,
  351, 193

\bibitem[{{Weisskopf} {et~al.}(2016{\natexlab{a}}){Weisskopf}, {Ramsey},
  {O'Dell}, {Tennant}, {Elsner}, {Soffitta}, {Bellazzini}, {Costa},
  {Kolodziejczak}, {Kaspi}, {Muleri}, {Marshall}, {Matt}, \& {Romani}}]{wro16}
{Weisskopf}, M.~C., {Ramsey}, B., {O'Dell}, S., {et~al.} 2016{\natexlab{a}}, in
  \procspie, Vol. 9905, Space Telescopes and Instrumentation 2016: Ultraviolet
  to Gamma Ray, 990517

\bibitem[{{Weisskopf} {et~al.}(2016{\natexlab{b}}){Weisskopf}, {Ramsey},
  {O'Dell}, {Tennant}, {Elsner}, {Soffita}, {Bellazzini}, {Costa},
  {Kolodziejczak}, {Kaspi}, {Mulieri}, {Marshall}, {Matt}, {Romani}, \& {IXPE
  Team}}]{wro16b}
{Weisskopf}, M.~C., {Ramsey}, B., {O'Dell}, S.~L., {et~al.} 2016{\natexlab{b}},
  Results in Physics, 6, 1179

\bibitem[{{Zdziarski} {et~al.}(1996){Zdziarski}, {Johnson}, \&
  {Magdziarz}}]{zjm96}
{Zdziarski}, A.~A., {Johnson}, W.~N., \& {Magdziarz}, P. 1996, \mnras, 283, 193

\bibitem[{{{\.Z}ycki} {et~al.}(1999){{\.Z}ycki}, {Done}, \& {Smith}}]{zds99}
{{\.Z}ycki}, P.~T., {Done}, C., \& {Smith}, D.~A. 1999, \mnras, 309, 561

\end{thebibliography}

\end{document}